%% file: main.tex
\documentclass[final, 11pt,5p]{elsarticle}

\usepackage{amssymb}

\usepackage{booktabs}
\usepackage{multicol}
\usepackage{multirow}
\usepackage{prettyref}
\newrefformat{eqn}{Equation~\ref{#1}}
\newrefformat{fig}{Figure~\ref{#1}}
\newrefformat{tab}{Table~\ref{#1}}
\newrefformat{sec}{Section~\ref{#1}}
\newrefformat{lst}{Listing~\ref{#1}}

\usepackage{caption}
\usepackage{subcaption}
\graphicspath{ {./pictures/} }

\journal{ARXIV}

\begin{document}

\begin{frontmatter}



\title{EXSCALATE: An extreme-scale in-silico virtual screening platform to evaluate \\1 trillion compounds in 60 hours on 81 PFLOPS supercomputers}


\author[polimi]{Davide Gadioli}
\author[polimi]{Emanuele Vitali}
\author[cineca]{Federico Ficarelli}
\author[cineca]{Chiara Latini}
\author[dompe]{Candida Manelfi}
\author[dompe]{Carmine Talarico}
\author[polimi]{Cristina Silvano}
\author[leonardo]{Carlo Cavazzoni}
\author[polimi]{Gianluca Palermo}
\author[dompe]{Andrea Rosario Beccari}

\affiliation[polimi]{organization={Politecnico di Milano - Department of Electronics, Information and Bioengineering},
            city={Milano},
            country={Italy}}

\affiliation[cineca]{organization={CINECA - SuperComputing Applications and Innovation Department},
            city={Casalecchio di Reno},
            country={Italy}}

\affiliation[dompe]{organization={Domp\'e farmaceutici S.p.A. - EXSCALATE},
            city={Napoli},
            country={Italy}}
            
             \affiliation[leonardo]{organization={Leonardo SPA},
            city={Genova},
            country={Italy}}


\begin{abstract}
The social and economic impact of the COVID-19 pandemic demands the reduction of the time required to find a therapeutic cure.
In the contest of urgent computing, we re-designed the Exscalate molecular docking platform to benefit from heterogeneous computation nodes and to avoid scaling issues.
We deployed the Exscalate platform on two top European supercomputers (CINECA-Marconi100 and ENI-HPC5), with a combined computational power of $81$ PFLOPS, to evaluate the interaction between $70$ billions of small molecules and $15$ binding-sites of $12$ viral proteins of Sars-Cov2.
The experiment lasted $60$ hours and overall it performed a trillion of evaluations.
\end{abstract}


\end{frontmatter}

\input{sections/introduction}

\input{sections/realated}

\input{sections/platform}

\input{sections/bigrun}
\input{sections/conclusion}

\section*{Acknowledgements}
This research was conducted under the project “EXaSCale smArt pLatform Against paThogEns for Corona Virus—Exscalate4CoV” founded by the EU’s H2020-SC1-PHE-CORONAVIRUS-2020 call, grant N. 101003551

\bibliographystyle{bibliography_style} 
\bibliography{bibliography}





\end{document}

%% file: sections/introduction.tex
\section{Introduction}

Drug discovery is a long process that usually involve \textit{in-silico}, \textit{in-vitro}, and \textit{in-vivo} stages.
The outcome of this process is a molecule, named \textit{ligand}, that has the strongest interaction with at least one binding site of the target protein, also known as \textit{pocket}, that represents the disease.
Domain experts expect this interaction to lead to a beneficial effect.
Virtual screening is one of the early stages that aims to select a set of promising \textit{ligands} from a vast chemical library.
The complexity of this operation is due to the ligand and pocket flexibility: both of them can change shape when they interact.
Therefore, to estimate the interaction strength using a \textit{scoring function}, we also need to predict the displacement of their atoms using a \textit{docking} algorithm.
This problem is computationally heavy, and it is well known in literature \cite{Pagadala2017}.
Moreover, to increase the probability of finding promising candidates, we would like to increase the size of the chemical library as much as possible, exacerbating the complexity of the virtual screening.
Since the evaluation is \textit{in-silico}, we can design new molecules by simulating known chemical reactions.
Therefore the chemical library size is limited only by the system's computational power.

In the context of urgent computing, where the time required to find a therapeutic cure should be as short as possible, we re-designed the Exscalate platform with the goal of virtual screening as many ligands as possible in a given time budget.
To maximize the throughput of the docking platform, we target \textit{High-Performance Computing (HPC)} supercomputers since their design maximizes the number of arithmetic operations per second \textit{Flop/s}, using double-precision floating-point numbers.
Indeed, the TOP500 list \cite{top500} ranks all the HPC supercomputers worldwide according to their throughput.
When we focus on the top five supercomputers, we can notice how four of them have heterogeneous nodes that heavily rely on accelerators, typically \textit{GPU}s.
Thus, we need to hinge on the node's accelerators and efficiently scale up to the available nodes to use all the computation power of the target machine.
Even if we focus on the software level, there are multiple well-known issues \cite{ashby2010opportunities,thakur2010mpi}.
How to efficiently use accelerators, how to transfer data within a node to feed the accelerators, how to move data from storage devices to the machine's node and vice versa, how to minimize communications between nodes and synchronizations between processes, and how to improve resilience to reduce the impact of faults in the time-to-solution, are the most representative issues.

\begin{figure*}[t!]
    \centering
    \includegraphics[width=\textwidth]{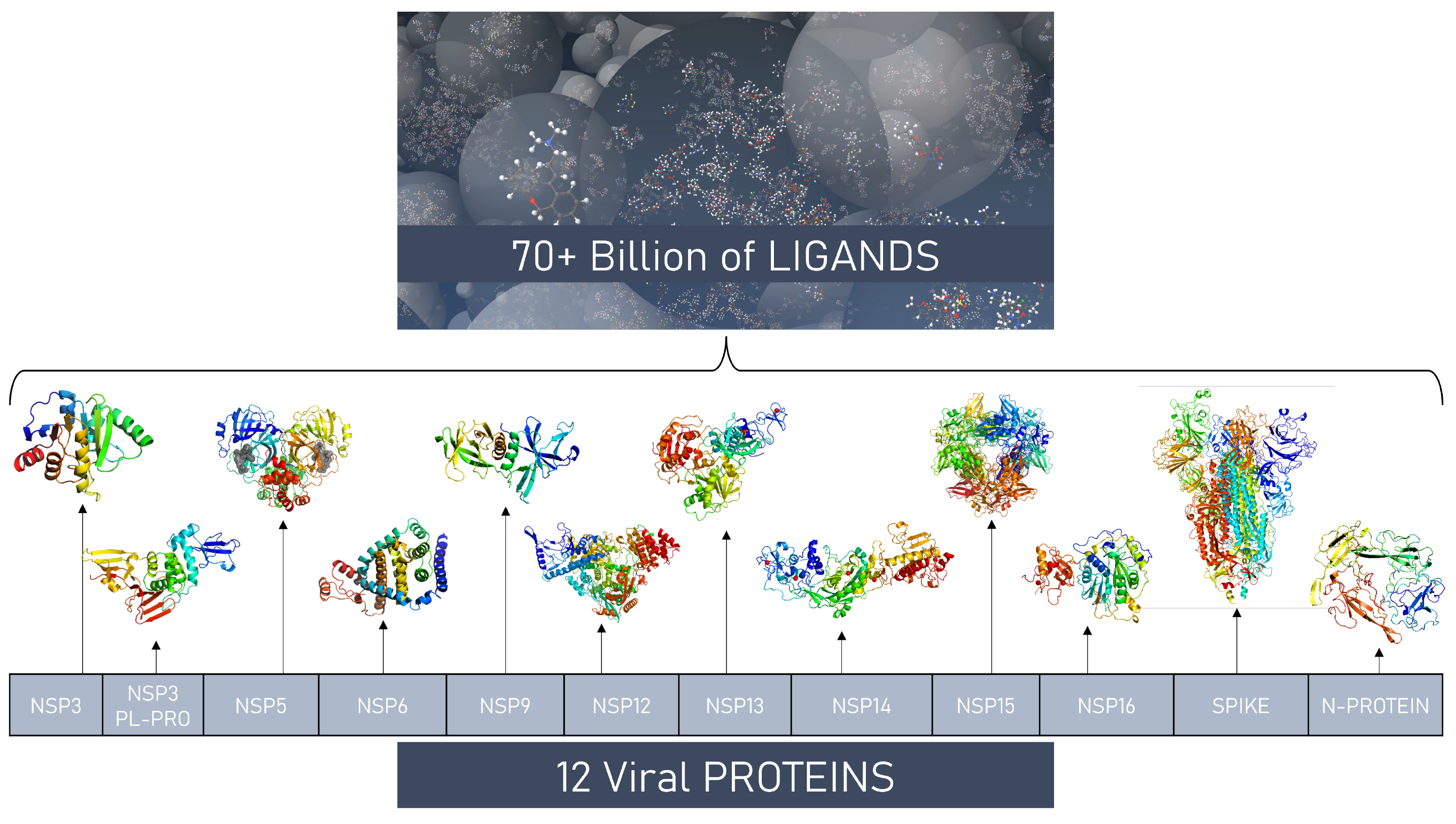}
    \caption{Schematic representation of the dataset used for the EXSCALATE4CoV virtual screening experiment.}
    \label{fig:1TD}
\end{figure*}

\begin{table}
    \centering
    \begin{tabular}{lr}
    \toprule
    \textbf{Protein} & \textbf{PDB code}\\
    \midrule
3CL protease (NSP5) & 6LU7 \\
N-protein & 6VYO \\
NSP3 & 6W02 \\
NSP6 & De novo model \\
NSP9 & 6W4B \\
NSP12 & 7BV2 \\
NSP13 & 6XEZ \\
NSP14 & Homology Model \\
NSP15 & 6W01 \\
NSP16 & 6W4H \\
PL protease & 6W9C \\
Spike-ACE2 & 6M0J \\
    \bottomrule
    \hline
    \end{tabular}
    \caption{\label{tab:targetPDB} The 3D targets used in the molecular docking experiments. A target might have different pockets.}
\end{table}

In the context of the EXSCALATE4CoV European project\footnote{https://www.exscalate4cov.eu/}, that targets to find new potential drugs against COVID19 pandemic \cite{COVID19}, we deployed the Exscalate platform in two HPC machines with a combined throughput of $81$ PFLOPS, to rank a chemical library of more than $70$ billion ligands against $15$ binding-sites of $12$ viral proteins of Sars-Cov2 (Figure \ref{fig:1TD}).
The crystal structures of the main functional units of SARS-CoV-2 proteome were obtained from the Protein Data Bank; \prettyref{tab:targetPDB} reports the list of the proteins analysed, with the corresponding PDB code.
Homology models of the proteins for which the crystal structure is not available were generated and used.
Overall, the experiment lasted $60$ hours and it performed a trillion of docking operations, becoming the largest virtual screening campaign up this moment.
The knowledge generated by this experiment is publicly released through the MEDIATE website\footnote{https://mediate.exscalate4cov.eu}.

The remainder of the paper is organized as follows: \prettyref{sec:related} briefly describes the most related application that can perform a virtual screen.
\prettyref{sec:platform} describes the Exscalate platform, highlighting the design choices that led to the performances reported in \prettyref{sec:bigrun}, where we deployed the application in two HPC machines.
Finally, \prettyref{sec:conclusion} concludes the paper.

%% file: sections/realated.tex
\section{Related works}
\label{sec:related}

A Molecular docking application can serve different purposes, from virtual screening to accurate simulations.
For this reason, we have a wide spectrum of algorithms and approaches\cite{Biesiada2011,Yuriev2015,Pagadala2017} that covers the performance-accuracy curve trade-off.
Since we use molecular docking to select the most promising candidates, we are interested in fast approaches such as ICM \cite{neves2012docking}, PSOVina \cite{Ng2015}, or EUDOCK \cite{Pang2010}.
However, fewer can use accelerators \cite{Thomsen2006}, which account for the majority of the computation power in the target supercomputers.

AutoDock \cite{Morris2009} is the most related work, since it has been ported in CUDA (AutoDock-GPU \cite{legrand2020gpu}) and deployed on the Summit supercomputer\cite{summit}, where they docked over one billion molecules on two SARS-CoV-2 proteins in less than two days \cite{glaser2021high}.
They hinge on the Summit's NVMe local storage to dock batches of ligands in the target pocket and to store the intermediate results.
In particular, AutoDock-GPU uses OpenMP to implement a threaded-based pipeline, where each thread reads ligands from the file, launches the CUDA kernels, waits for their completion, and it writes back the results.
Since most docking algorithms use a fast but approximated scoring function to drive the estimation of the 3D pose of a ligand, it is common to re-score the most promising ones with a more accurate scoring function.
They use a custom CUDA version of RFScore-VS \cite{wojcikowski2017performance} to perform such task and BlazingSQL \cite{blazingSQL} for computing statistics and selecting the top scoring ligands.
To orchestrate the workflow they use FireWorks \cite{jain2015fireworks} from an external cluster to ensure a consistent state in presence of faults in the computations nodes.

The approach that we followed to design the Exscalate platform differs in several ways.
We use a monolithic application to dock and re-score the ligands, using MPI \cite{message2015mpi} to scale out and \textit{C\texttt{+}\texttt{+}11} threads to scale up.
The proposed solution can reach a high throughput without relying on the node's local storage, which is not available in the target HPC systems.
Moreover, we envelop the application in a more complex workflow that compensates for its weak points.
\prettyref{sec:platform} will describe the platform in more detail.
Furthermore, since our docking algorithm is deterministic, we can store only the molecule's structure, using the SMILES format \cite{weininger1988smiles}, and the best score that we found.
Then, we can re-generate the 3D displacement of the atoms on demand.

%% file: sections/platform.tex
\section{Exscalate platform}
\label{sec:platform}

This section describes the Exscalate platform from a software engineering point of view.
\prettyref{sec:bigrun} provides more detail on how we tailored the platform for the experiment.

\subsection{The dock and score algorithm}
\label{sec:dock_and_score}

\begin{figure*}[t]
    \centering
    \begin{subfigure}[b]{0.49\textwidth}
        \centering
        \includegraphics[width=\textwidth]{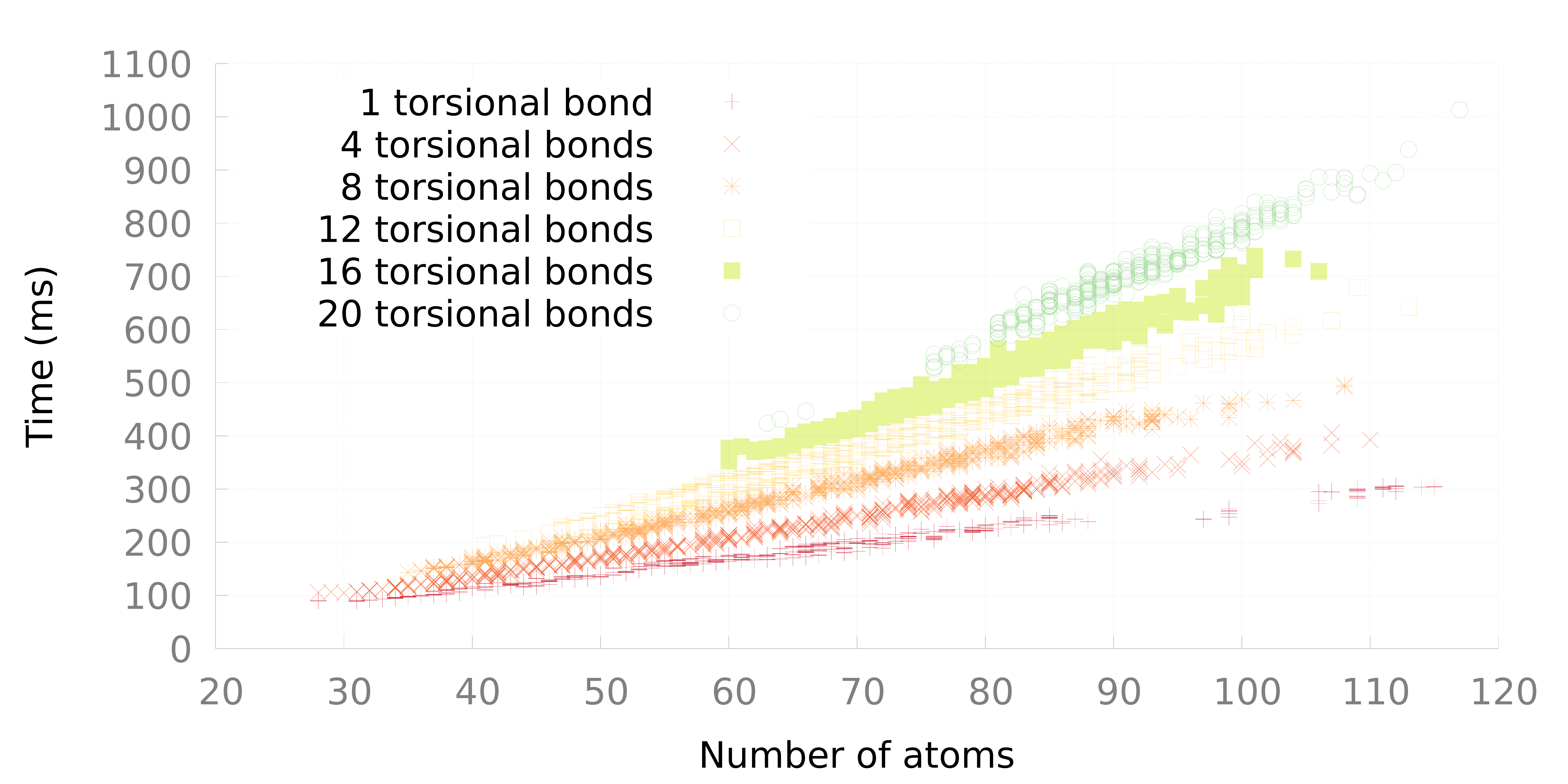}
        \caption{\textit{C\texttt{+}\texttt{+}} implementation on CPU}
        \label{fig:dock_time_cpu}
    \end{subfigure}
    \hfill
    \begin{subfigure}[b]{0.49\textwidth}
        \centering
        \includegraphics[width=\textwidth]{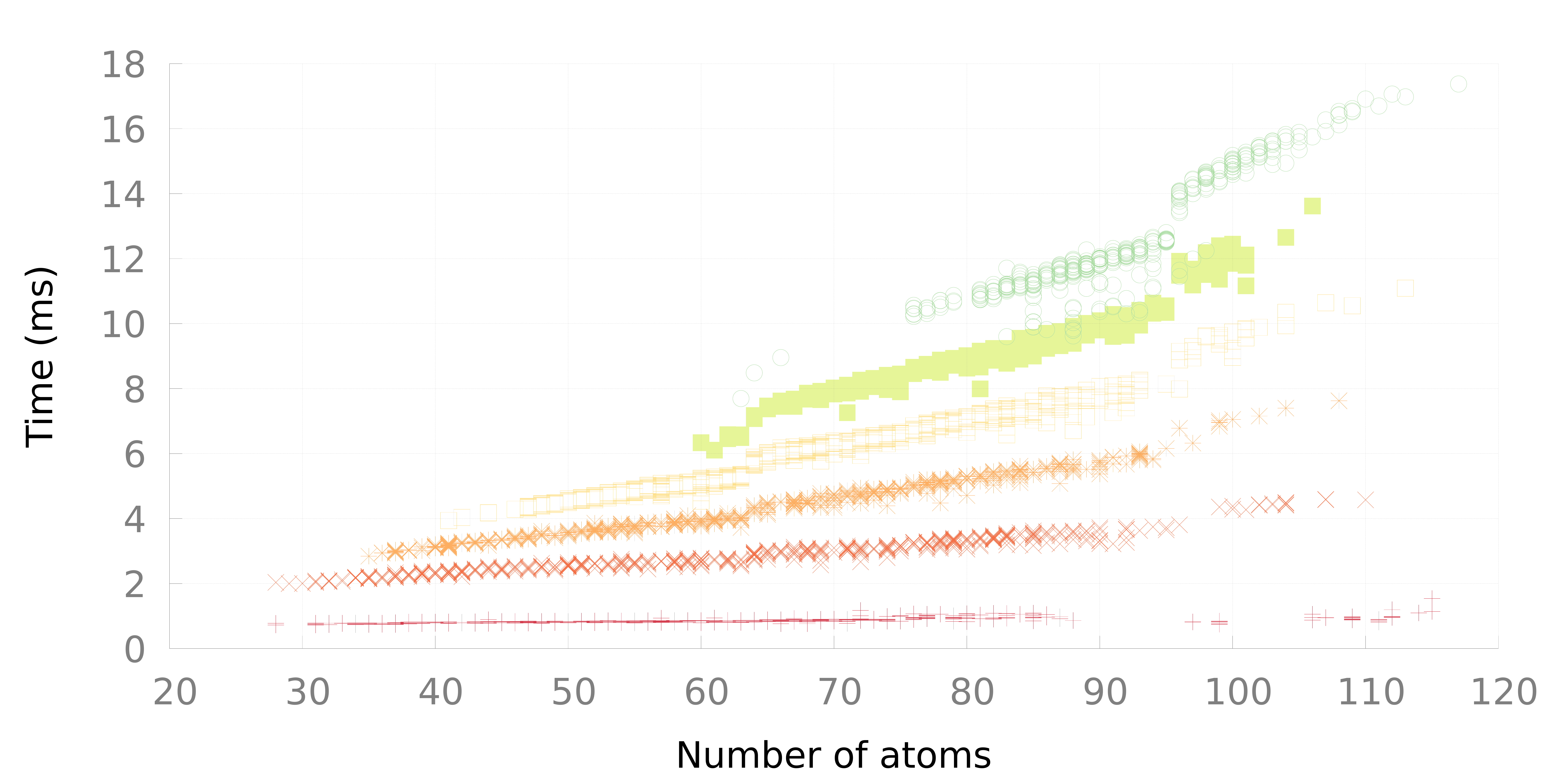}
        \caption{CUDA implementation on GPU}
        \label{fig:dock_time_gpu}
    \end{subfigure}
    \caption{Time required to dock and score a ligand by varying the number of atoms and torsional bonds. The C\texttt{+}\texttt{+} implementation use a single core IBM $8335$-GTG $2.6$ GHz. The CUDA implementation use a single NVIDIA V$100$.}
    \label{fig:dock_time}
\end{figure*}

The final output of the algorithm is an estimation of the bond strength between a given ligand and the binding site of the target protein.
In the virtual screening context, it is common to reduce the problem's complexity by using heuristics and empirical rules instead of performing a molecular dynamic simulation \cite{cheng2012structure}.
One implication of this choice is that the numeric score of a ligand is strongly correlated by the given 3D displacement of its atoms, which is not trivial to compute due to the high number of degrees of freedom involved in the operation.
Besides the six degrees of freedom derived by rotating and translating a rigid object in a 3D space, we must consider the ligand flexibility.
A subset of the ligand's bonds, named \textit{torsional bonds} \cite{veber2002molecular}, partition the ligand's atoms in two disjoint set, that can rotate along the bond's axis, changing the ligand's shape.
A small molecule can have tens of torsional bonds.

The algorithm that we use in Exscalate to score a ligand is composed of four steps.
The first step is a ligand pre-processing that flattens the ligand by rotating the torsional bonds to maximize the sum of the internal distances between all the molecule atoms.
This computation is protein independent.
The second step docks the ligand inside the binding site of the target protein by using a greedy optimization algorithm with multiple restarts.
The scoring function that we use to drive the docking considers only geometrical steric effects.
We take into account the ligand's flexibility, but we consider the pocket as a rigid body \cite{10.1145/3235830.3235835}.
In the experiment, we evaluated $256$ different initial poses for each ligand.
The third step sorts the generated poses to select only a few to re-score using the LiGen chemical scoring function \cite{beato2013use} in the fourth step.
In particular, we cluster the generated poses using a root mean square deviation (RMSD) of $3A$ as the threshold to deem two poses as similar.
Then we sort them to have in the first places the top-scoring pose for each cluster, then all the others; sorted according to the geometrical scoring function.
In the experiment, we scored only the top $30$ poses for each ligand.
The score of the ligand is the score of the best pose that we found.

We implemented the algorithm in \textit{C\texttt{+}\texttt{+}17} to target CPUs, while we implmented an OpenACC and CUDA implementation to target accelerators \cite{vitali2019exploiting}.
The CUDA implementation is the fastest when we target NVIDIA GPUs.
The algorithm's asymptotic complexity is $O(n \cdot m)$, where $n$ is the number of atoms and $m$ is the number of torsional bonds.
We omit features of the target docking site since it is constant during the docking application lifetime.
\prettyref{fig:dock_time} measures the time required to dock and score a ligand on a Marconi100 node\cite{m100}, by varying the implementation, the ligand's number of atoms, and torsional bonds.
While the C\texttt{+}\texttt{+} implementation's performance (\prettyref{fig:dock_time_cpu}) is expected, the CUDA one (\prettyref{fig:dock_time_gpu})) is less related with number of atoms.
This is due to the fact that we can use hardware parallelism to elaborate them, while we need to process the torsional bonds serially to preserve the molecule geometry.
Moreover, since we organized the elaboration in bundles of $32$ atoms, we have a steep increase in the docking time when we need to process more atoms; for example after $64$ and $96$ atoms.

We can notice in both implementations how the docking time is heavily input dependent, where the difference between the fastest and slowest class of ligands is more than one order of magnitude.

\subsection{Exscalate high-throughput docking application}
\label{sec:docker}

The only information required to dock and score a ligand in the target binding site is their description.
Thus, the virtual screening process is an embarrassingly parallel problem.
However, it is of paramount importance to design how the data can be read from the storage, transferred to the accelerator, and written back to the storage.

\begin{figure}[t]
    \centering
    \includegraphics[width=0.4\textwidth]{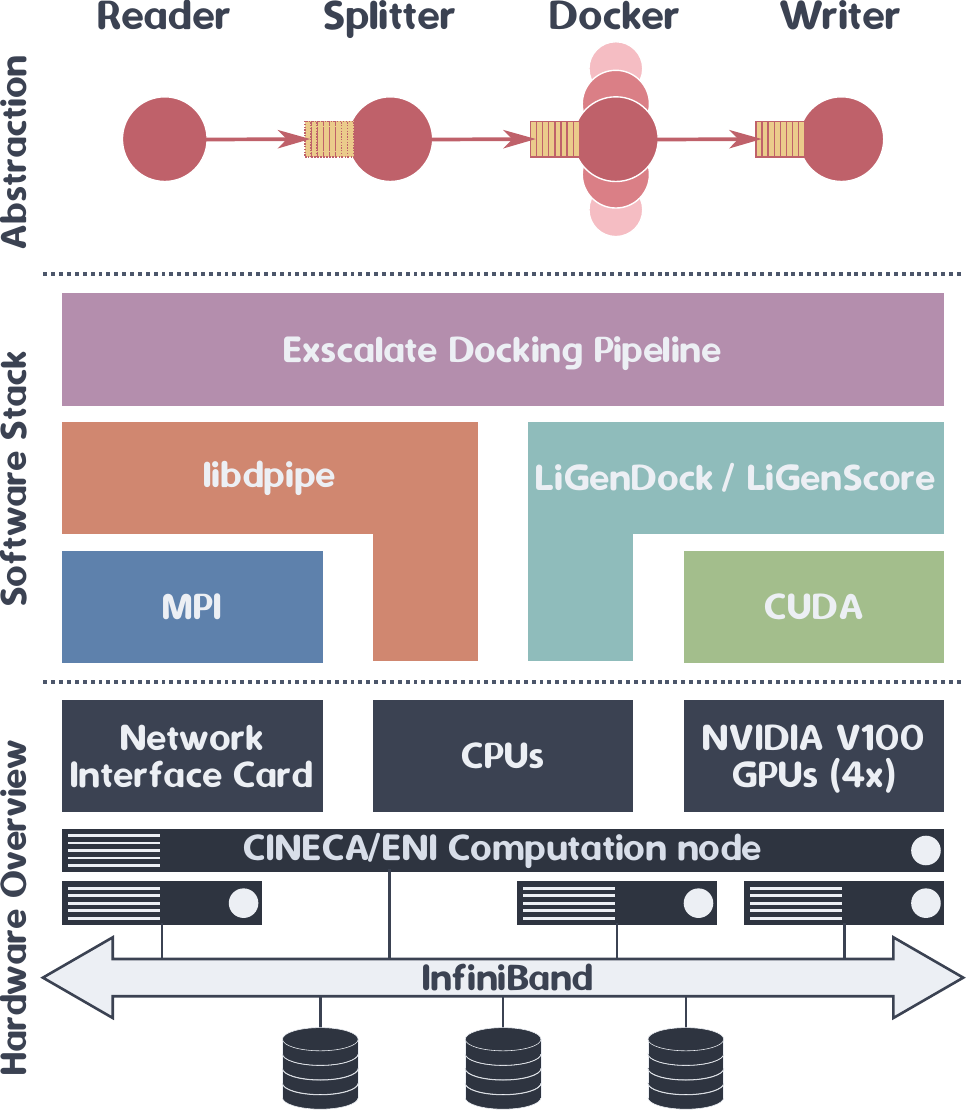}
    \caption{Overview of the Exscalate docking application by varying the abstraction level.}
    \label{fig:docker_ht}
\end{figure}

\prettyref{fig:docker_ht} shows an overview of the application abstraction and software stack of the Exscalate high-throughput docking application.
We have chosen to write an MPI application that implements an asynchronous pipeline.
In particular, we want to execute a single process for each node available.
Then, each process spawns a pipeline to carry out the elaboration using all the computation resources of its node.
We use a dedicated thread for each stage of the pipeline.
Moreover, each stage may have a thread-safe queue that stores input data.

The first stage is the \textit{reader}, which reads from the actual file that represents the chemical library a chunk of data that it enqueues in the \textit{splitter}'s queue.
The splitter stage inspects each chunk of data to separate all the ligand's descriptions that contain.
Then it enqueues each ligand description in the \textit{docker}'s queue.
In the experiment, we describe a ligand using a custom binary format derived from the TRIPOS Mol2 format, described in more detail in \prettyref{sec:bigrun}.
The docker stage dequeues a ligand description, it constructs the related data structures, performs the dock and score steps described in \prettyref{sec:dock_and_score}, and it enqueues the ligand's score in the \textit{writer}'s queue.
The writer stage dequeues the ligand score and accumulates in an internal buffer the related output, which is the ligand's SMILES representation and its score value in a CSV-like fashion.
When the accumulation buffer is full, the writer stage initiates the writing procedure.

The docker stage is the only one that can be composed of several threads that operate on the same queues to enable work-stealing.
Moreover, it is possible to use different algorithm implementations, such as CUDA and \textit{C\texttt{+}\texttt{+}}, to leverage the node's heterogeneity.
We refer to any docker thread as \textit{worker}.
All the workers that use the CUDA implementation are named \textit{CUDA workers}, while the ones that use the \textit{C\texttt{+}\texttt{+}} implementation are named \textit{CPP workers}.
Even if a single CUDA worker is tied to a single GPU, it is possible to have multiple CUDA workers tied to the same GPU. 
We consider the target binding site constant during the elaboration.
Therefore, each process will fetch the related information once at the beginning of the execution.
Each algorithm implementation is free to store the pocket data structures in the most appropriate memory location during its initialization.
In particular, the \textit{C\texttt{+}\texttt{+}} implementation uses constant static memory, while the CUDA implementation uses texture memory.

The Exscalate Docking Pipeline library contains all the application's stages implementations.
To parallelize the computation it uses the high-level interfaces for MPI and \textit{C\texttt{+}\texttt{+}} threads provided by libdpipe.
The LiGenDock/LiGenScore libraries implement the domain-specific functional concerns.

\begin{figure}[t]
    \centering
    \includegraphics[width=0.48\textwidth]{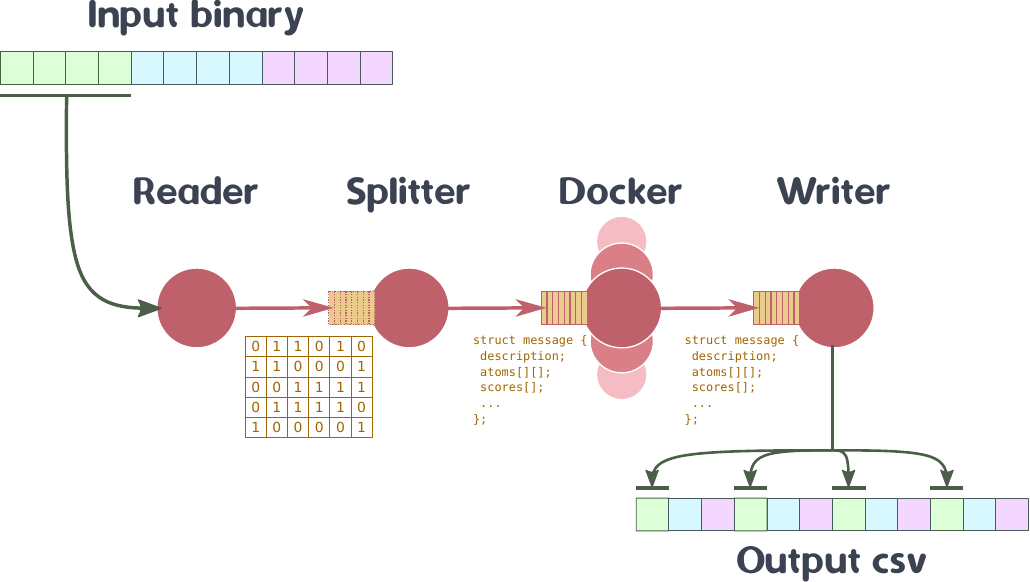}
    \caption{Example of I/O synchronization with three Application instances, represented by different colors, that read the input ligands and store the results.}
    \label{fig:docker_io}
\end{figure}

Even if the problem is embarrassing parallel, we need to synchronize all the application's instances using MPI when we perform I/O operations.
\prettyref{fig:docker_io} depicts an example of I/O coordination with three MPI processes, represented by different colors.
Since the computation pipeline is the same for all the processes, we depict only one MPI process pipeline.

To distribute the computation workload among the MPI processes, we split the input file in even slabs according to the file size and the number of MPI processes.
Since the size of a molecule description depends on the ligand's properties, such as the number of atoms, it seldom happens that a slab starts with the beginning of a ligand description and it stops with the ending of a molecule description.
We use the convention that each process elaborates all the ligands whose description begins between the slab start and stop.
The last ligand description may end after the slab stop.

On the main hand, we are using a very I/O-friendly access pattern because we read a file sequentially.
On the other hand, the static data partition negates work-stealing among MPI processes.
Therefore, the application throughput is equal to the throughput of the slowest process.
The frequency at which each process reads from the input file depends on the pipeline throughput.

The writer stage uses collective I/O operations to coalesce writing requests together before writing to the storage.
Indeed, the user can configure the number of processes that issue I/O operations to reduce the pressure on the file system.
This access pattern is I/O-friendly because all the writing operations are parallel and sequential.

\subsection{Exscalate workflow}
\label{sec:workflow}

In principle, it is possible to store all the ligands that we would like to dock in a single file and to deploy the docking application on the whole machine.
However, this approach has several drawbacks.
The main concern is fault resiliency.
The default action to respond to a fault in an MPI communicator, for example after a node failure, is to terminate all the processes \cite{message2015mpi}, which can lead to losing a significant amount of computation effort.
This is a well known problem in literature \cite{snir2014addressing,bland2013post,Rocco2021}.
Another concern lies in the application performance.
\prettyref{fig:dock_time} shows how docking and scoring a large and complex ligand required much more time than a small and simple ligand.
Therefore, we have a significant imbalance between the MPI processes if all the ligands with many atoms and torsional bonds are close.

\begin{figure*}[t]
    \centering
    \includegraphics[width=\textwidth]{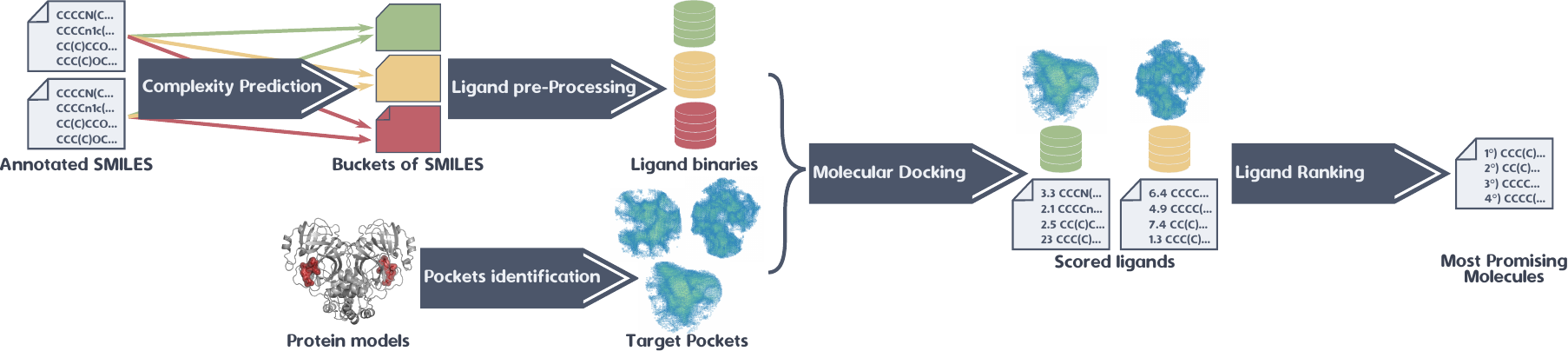}
    \caption{Exscalate workflow, from the input (ligand's chemical library and the protein models) on the left; to the final outcome (most promising set of molecules) on the right.}
    \label{fig:workflow}
\end{figure*}

The Exscalate workflow address these issues with a pre-processing phase on the chemical library to have a relatively small number of jobs that can run in parallel using a plain job array to coordinate the execution, such as the one provided by SLURM \cite{10.1007/10968987_3} or PBS \cite{pbs}.
\prettyref{fig:workflow} depicts the Exscalate workflow which requires two different kinds of input from the domain knowledge.
On one hand, we require the binding sites of the target proteins.
How to obtain them is a complex procedure that we consider outside of the scope of this article \cite{ijms21145152}.
On the other hand, we require the chemical library of molecules that we want to evaluate.
It is possible to represent a molecule in a wide range of formats according to the amount of information that we want to store.
In our case, we assume that the chemical library is stored using the SMILES format \cite{doi:10.1021/ci00057a005}, which can encode a molecule in a single string that contains only the structure of the molecule (ignoring the hydrogen) since it is the most compact.

The next step in the ligand pre-processing is to broadly classify them in buckets according to their expected execution time, to reduce as much as possible the imbalance during the computation.
As shown in \prettyref{fig:dock_time}, the number of torsional bonds and atoms seem good predictors.
However, it is not trivial to extract these properties from the SMILES representation.
For this reason, we trained a model that predicts the execution time given properties that are more accessible at this point of the workflow: the number of heavy atoms, the number of rings, and the number of chains.
We also consider interactions between them.
We use a decision tree model with a maximum depth of $16$ to predict the ligand's execution time.

After the ligands classification according to their complexity, we can perform the pre-processing.
In particular, for each ligand we add the hydrogen atoms, we generate the initial displacement of its atoms in the 3D space, and we unfold the molecule (\prettyref{sec:dock_and_score}).
This elaboration is required once and it can be re-reused in all the virtual screening campaigns.
In addition to the ligand pre-processing, we also need to set the size of the ligand binary files that we will use in the docking phase.

Finally, once we have the target binding sites and the ligand binaries, we can perform the virtual screening campaign.
The idea is to launch the docking application on all the ligand files, one pocket at a time.
The output of the virtual screening is the ranking of the chemical library against each docking site.
When domain experts selected the ligands that have a strong interaction with multiple docking sites or proteins, it is possible to re-create the 3D displacement of the ligand's atoms on demand.
For this reason, we can store only the structure of the molecule, using the SMILES notation.

%% file: sections/bigrun.tex
\section{Experimental results}
\label{sec:bigrun}


We validate the Exscalate platform by performing a virtual screening campaign over $70$ million ligands against $15$ binding sites of $12$ viral proteins of Sars-Cov2.
We deployed the Exscalate platform on HPC5 at ENI\cite{hpc5} and Marconi100 at CINECA\cite{m100}.
A Marconi100 node is equipped with $32$ IBM POWER9 AC922 cores ($128$ hardware threads) and $4$ NVIDIA V100 GPU, with NVLINK $2.0$.
The computation node of HPC5 is very similar since it also uses $4$ NVIDIA V100 GPUs, but it relies on Intel Xeon Gold 6252 24C as CPU ($24$ cores and $48$ hardware threads) and it uses NVLINK only for the GPU interconnection.
This section reports the extra-functional concerns of the experiment.

\subsection{Evaluating the storage requirements}

\begin{figure*}[t]
    \centering
    \begin{subfigure}[b]{0.49\textwidth}
        \centering
        \includegraphics[width=\textwidth]{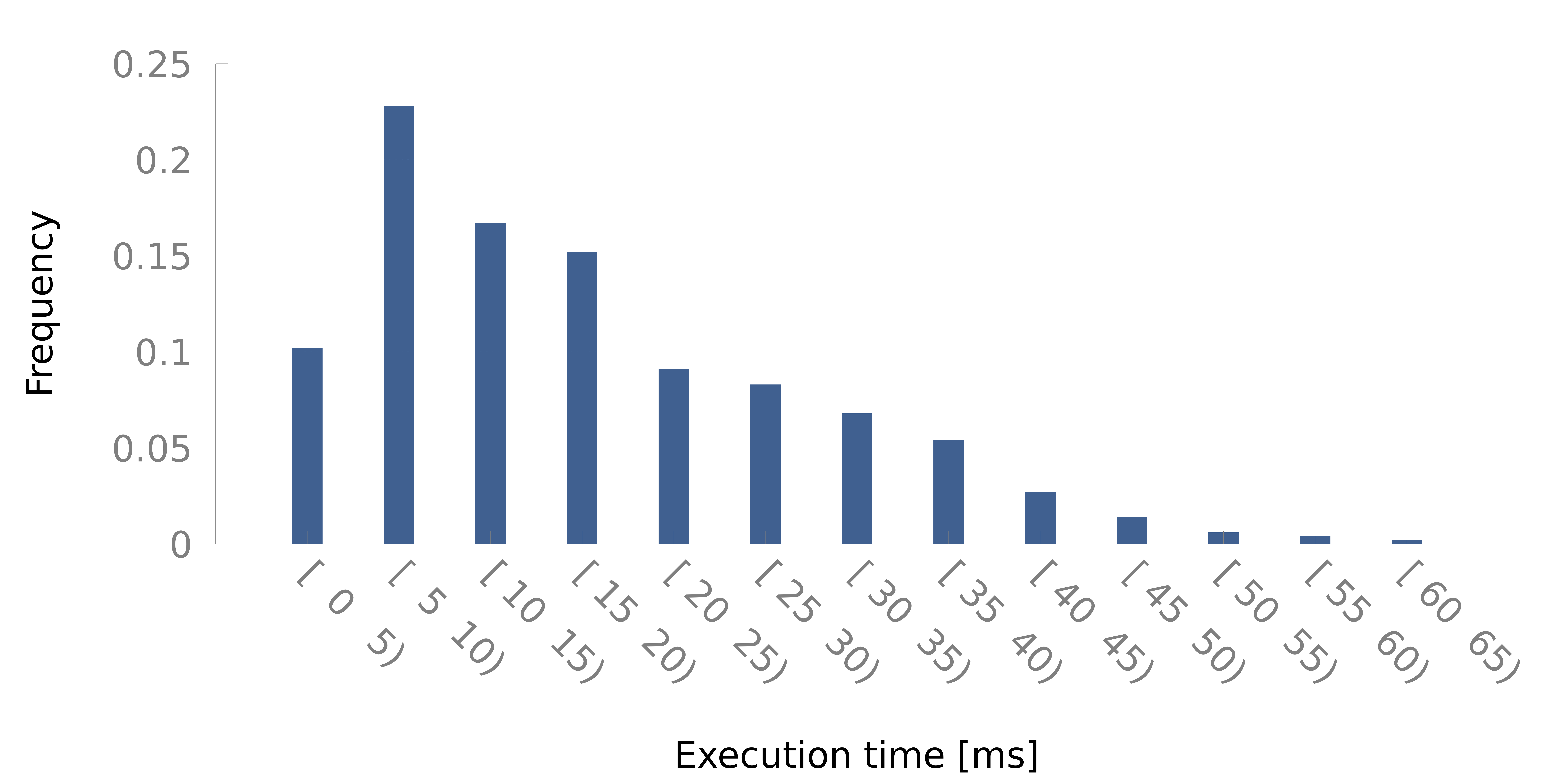}
        \caption{Measured docking time}
        \label{fig:validation_time}
    \end{subfigure}
    \hfill
    \begin{subfigure}[b]{0.49\textwidth}
        \centering
        \includegraphics[width=\textwidth]{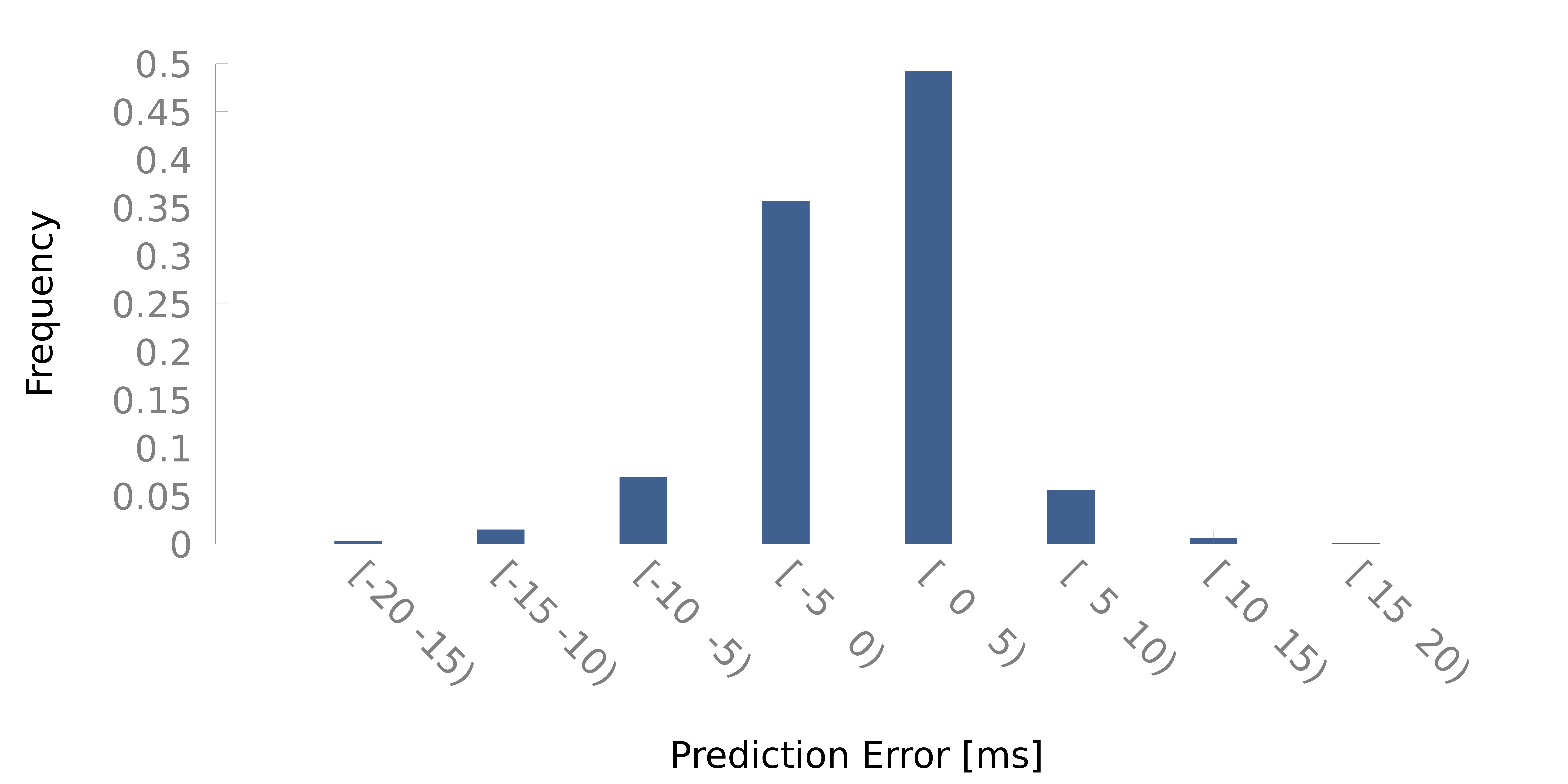}
        \caption{Prediction error}
        \label{fig:validation_error}
    \end{subfigure}
    \caption{Frequency distribution of the measured docking time, using the CUDA implementation, and its prediction error. We discared values with a frequency lower than $0.001$ for conciseness purposes.}
    \label{fig:validation}
\end{figure*}

One of the main concerns in HPC systems is the storage.
When scaling an experiment to the scale of a trillion docking operation, it requires us to evaluate in detail what we want to read and write, giving attention to the format.

To perform the virtual screening, we need information about the binding sites of the target proteins and the chemical library of ligands that we want to analyze.
The former is not an issue since it requires total storage of $29$MB and the information needed is read once when the application starts.
The latter needs more careful consideration.
Domain experts use to work with SMILES format to represent a ligand since it is the most compact.
In fact, the chemical library evaluated in the experiment encoded in the SMILES format requires a total of $3.3$TB.
However, the docking application requires a richer description of the molecule, as described in \prettyref{sec:workflow}.
The most widely used format to store the required information is the TRIPOS Mol2, which is encoded in ASCII characters and focuses on readability rather than efficiency.
For this reason, we use a custom binary format that stores only the information required by the docking application, such as the atom's position, type, and bonds.
By comparing the size of the same molecules, the Mol2 format requires $5-6$X more space with respect to the binary format.
Nonetheless, the whole binary chemical library for the experiment requires $59$TB of storage.

Storing all the docked poses is unfeasible since we are targeting $15$ binding site and we re-score $30$ alternative pose for each input ligand because it would require $26$PB of storage.
For this reason, we store only the SMILES of the molecule and its best score in a CSV-like file.
Then, we can re-generate the docked pose on demand since the docking algorithm is deterministic.
The size of the final output is $7.3$TB of data.

\subsection{Predicting the ligand complexity}
\label{sec:validation}

\begin{figure}[t]
    \centering
    \includegraphics[width=0.49\textwidth]{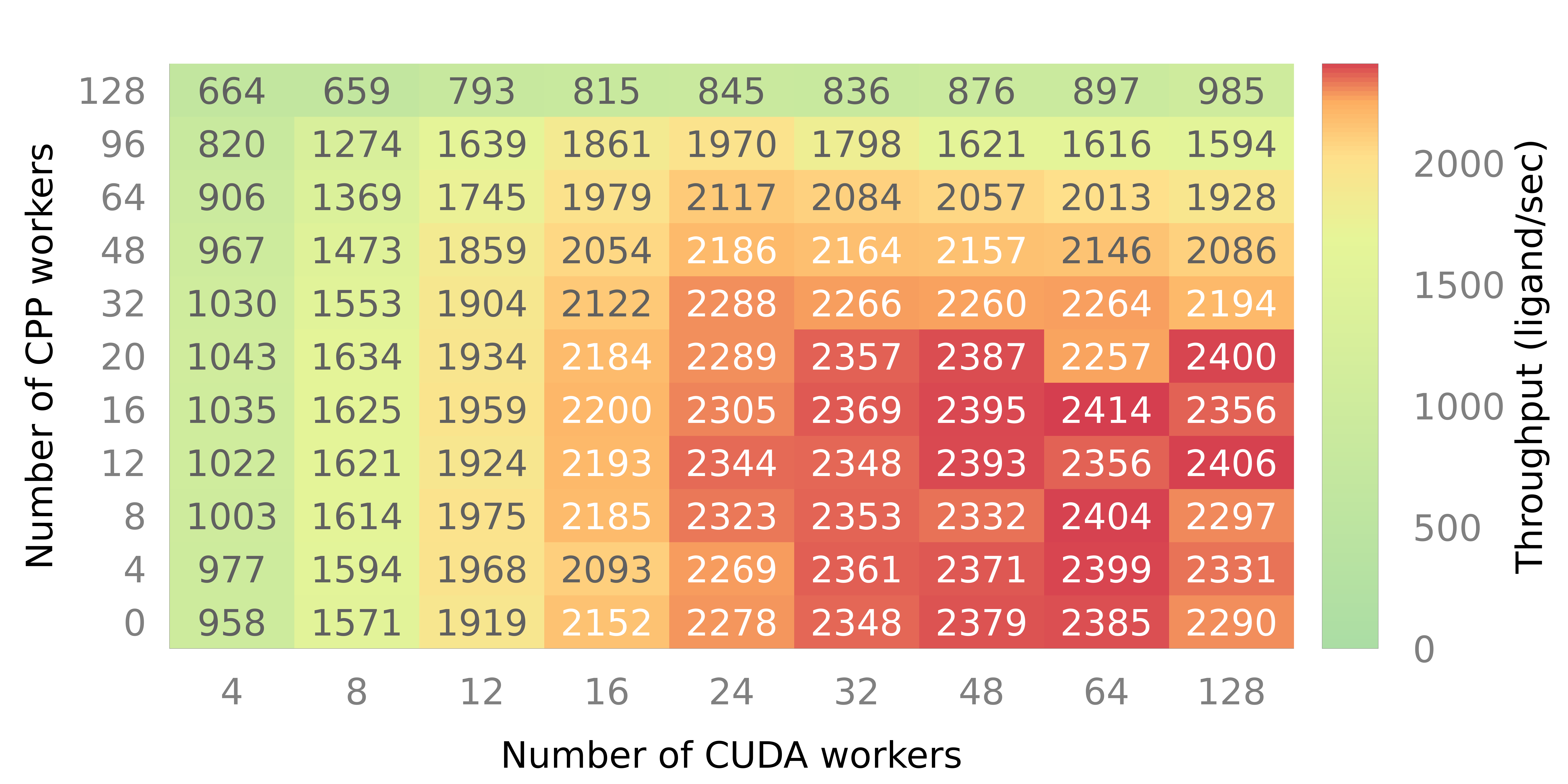}
    \caption{Throguhput of the docking application in terms of ligands per seconds, by varying the number of CPP and CUDA workers.}
    \label{fig:throughput}
\end{figure}

\begin{table*}
    \centering
    \begin{tabular}{lccr}
    \toprule
    Binding site & Thr (ligands/sec/node) & Thr (ligands/sec) & HPC machine \\
    \midrule
    PLPRO	    &2496	    &1996800            &M100 \\
    SPIKEACE	&2498	    &1998400            &M100 \\
    NS12thumb	&2499	    &1999200            &M100 \\
    NS13palm	&2486	    &1988800            &M100 \\
    3CL	        &2427	    &1941600            &M100 \\
    NSP13allo	&2498	    &1998400            &M100 \\
    Nprot	    &2010       &3015000            &HPC5 \\
    NSP16	    &1980	    &2970000            &HPC5 \\
    NSP3	    &1969	    &2953500            &HPC5 \\
    NSP6	    &1985	    &2977500            &HPC5 \\
    NSP12ortho	&2001	    &3001500            &HPC5 \\
    NSP14	    &1965	    &2947500            &HPC5 \\
    NSP9	    &1996	    &2994000            &HPC5 \\
    NSP15	    &1990	    &2985000            &HPC5 \\
    NSP13ortho	&2454/1987	&1963200/2980500    &M100/HPC5 \\
    \bottomrule
    \hline
    \end{tabular}
    \caption{\label{tab:throughput}The throughput reached per node and per machine for each binding site evaluated in the experiment.The NSP13ortho binding site has been partially computed in both machines.}
\end{table*}

The main shortcoming of the docking application is its inability to perform work-stealing among different nodes, potentially leading to an imbalance in the execution.
For this reason the workflow clusters the input chemical library according to the expected time required to elaborate a ligand, by using features that are trivially accessible from the SMILES format.

\prettyref{fig:validation} shows the experimental campaign that we used to train a decision tree regressor \cite{scikit-learn} to predict the docking time using the number of heavy atoms, the number of rings, the number of chains, and the interactions between them.
\prettyref{fig:validation_time} shows the measured execution time of a dataset with $21$ million of ligands with a different number of atoms and torsional bonds.
We use the $80\%$ of the data to train the model, while the remaining data are used to compute the prediction error reported in \prettyref{fig:validation_error}.
The model has a negligible mean error ($-0.00088$ ms), with a standard deviation of $3.81$ ms.

Even if the average error is close to zero, we can notice how the standard deviation suggests that we do make an error when we predict the docking time of a given ligand.
In the experiment, we cluster the ligands in buckets of $10$ ms to account for this variability.
Since the high-throughput docking aims at avoiding imbalance in the computation, we are interested in the average behavior.

\subsection{Exploiting a node heterogeneity}

The availability of multiple implementations of the dock and score algorithm grants access to heterogeneous resources.
However, the relationship between the number of CUDA and CPU workers (\prettyref{sec:docker}) and the application throughput is not trivial.
\prettyref{fig:throughput} shows the application throughput in terms of docked ligands per second, by varying the number of CUDA and CPP workers, when we deploy the application on a Marconi100 node, which has $32$ IBM POWER9 AC922 cores ($128$ hardware threads) and $4$ NVIDIA V100 GPU.
The application binds each CUDA worker to a single GPU in a round-robin fashion.
For example, when we use $24$ CUDA workers, we have $6$ threads that feed data and retrieve the results for each GPU in the node.

From the throughput, we can notice how the application reaches the peak performance for a high number of CUDA workers.
Moreover, when we increase the number of CPU workers to match the number of hardware threads, we harm the application performance.
This behavior implies that, in our case study, it is better to use CPUs to support accelerators and I/O operations rather than contribute to the elaboration itself.
Furthermore, to benefit most from a GPU it is not enough to use a single CUDA worker.
We expect this result because the CUDA worker needs to parse the ligand description and initialize the related data structures before launching any CUDA kernel.
Thus, by using more CUDA workers we can hide these overheads and fully utilize the GPU.

To perform this analysis we used the dataset ``Commercial Compound MW$<$330'' from the MEDIATE website.
It is composed of $5$ milion of small molecules and it is publicly available.

\subsection{Scaling on the target HPC machine}

To overcome the limitations of using a single MPI application that runs on the whole supercomputer, we run several instances on different portions of the input data.
We divided the input set in {\raise.17ex\hbox{$\scriptstyle\mathtt{\sim}$}}$3400$ jobs, where each job is composed of $32$ MPI processes that last for about $5$ minutes and targets a single binding site.
For this experiment, we decided to evaluate the binding sites sequentially.

\prettyref{tab:throughput} reports for each binding site the average throughput of a node and the whole machine.
In particular, to compute the throughput of a node, we sampled the application log files to get the average throughput per MPI process, which is equal to the node performance and then computed the average.
To compute the machine throughput,  we divided the time to solution of the experiment, i.e. the wall time required to dock the whole chemical library against the target docking site, by the number of ligands in the chemical library.
Thus, it takes into consideration all the overheads of the execution.
On average, the throughput of a single node is $2.2k$ ligands per second, while the combined throughput of the both supercomputers is $5M$ ligands per second.

The throughput per node that we measured while scaling to the whole machine is comparable to the one that we measured while tuning the number of CUDA and CPU workers.
Therefore, the Exscalate platform was able to exploit all the available resources.

%% file: sections/conclusion.tex
\section{Conclusion}
\label{sec:conclusion}

In the context of urgent computing, where we want to reduce the social and economic impact of a pandemic as much as possible, we re-designed the Exscalate molecular docking platform targeting HPC systems.
We use this platform to perform the largest virtual screening campaign against $15$ binding sites of $12$ viral proteins of Sars-Cov2.
The results accounts for 64TB of data representing the score of each of the 70+ billion of ligands on the target pockets. 
The set of most promising compounds filtered for each target has been made available on the MEDIATE portal\footnote{https://mediate.exscalate4cov.eu/data.html} to permit researchers around the world to start more detailed de-novo campaigns from a reduced set of compounds.

This document describes the Exscalate platform design from a software engineering point of view and reports its validation in the experiment.
In particular, the experiment shows how the docking application can hinge on a node's accelerators to carry out the computation while using the CPU mainly to support the computation; i.e. by synchronizing the I/O toward the file system and feeding the GPUs with data.
Moreover, we were able to scale over the two full HPC system, CINECA-Marconi100 and ENI-HPC5, that at the time of the experiment were the most powerful European supercomputers.